\documentclass[aps,prd,10pt,nofootinbib,twocolumn,eqsecnum,showpacs,showkeys,superscriptaddress,preprintnumbers,altaffilletter]{revtex4-1}
\usepackage{graphicx}
\usepackage{amsmath}
\usepackage{multirow}
\usepackage{hyperref}
\usepackage{soul}
\usepackage{graphicx}
\usepackage{dcolumn}
\usepackage{amssymb}
\usepackage{amsfonts}
\usepackage{amsbsy}
\usepackage{color}
\usepackage{rotating}
\usepackage[english]{babel}
\usepackage{multirow}
\usepackage{float}
\usepackage{orcidlink}

\usepackage[normalem]{ulem}

\newcommand{\be}{\begin{equation}}
\newcommand{\ee}{\end{equation}}
\newcommand{\bea}{\begin{eqnarray}}
\newcommand{\eea}{\end{eqnarray}}
\newcommand{\beal}{\begin{aligned}}
\newcommand{\eeal}{\end{aligned}}
\newcommand{\bi}{\begin{itemize}}
\newcommand{\ei}{\end{itemize}}

\newcommand{\pa}{\partial}

\hypersetup{
    colorlinks=true,
    linkcolor=blue,
    filecolor=magenta,
    urlcolor=cyan,
    citecolor=cyan
}

\begin{document}

\title{Gravitational lensing from clusters of galaxies to test Disformal Couplings Theories}

\author{Saboura Zamani\,\orcidlink{0009-0004-3201-9483}}
\email{saboura.zamani@phd.usz.edu.pl}
\affiliation{Institute of Physics, University of Szczecin, Wielkopolska 15, 70-451 Szczecin, Poland}
\author{Vincenzo Salzano\, \orcidlink{0000-0002-4905-1541}}
\email{vincenzo.salzano@usz.edu.pl}
\affiliation{Institute of Physics, University of Szczecin, Wielkopolska 15, 70-451 Szczecin, Poland}
\author{Dario Bettoni\, \orcidlink{0000-0002-0176-5537}}
\email{dbet@unileon.es}
\affiliation{Departamento de Matemáticas, Universidad de León,
Escuela de Ingenierías Industrial, Informática y Aeroespacial
Campus de Vegazana, s/n
24071 León}
\affiliation{IUFFyM, Universidad de Salamanca, E-37008 Salamanca, Spain}

\date{\today}

\begin{abstract}

In this study, we investigate the potential existence of a non-minimal coupling between dark matter and gravity using a compilation of galaxy clusters. We focus on the disformal scenario of a non-minimal model with an associated coupling length $L$. Within the Newtonian approximation, this model introduces a modification to the Poisson equation, characterized by a term proportional to $L^2 \nabla^2 \rho$, where $\rho$ represents the density of the DM field. We have tested the model by examining strong and weak gravitational lensing data available for a selection of 19 high-mass galaxy clusters observed by the CLASH survey.
We have employed a Markov Chain Monte Carlo code to explore the parameter space, and two different statistical approaches to analyse our results: a standard marginalisation and a profile distribution method. Notably, the profile distribution analysis helps out to bypass some volume-effects in the posterior distribution, and reveals lower Navarro--Frenk--White concentrations and masses in the non-minimal coupling model compared to general relativity case. We also found a nearly perfect correlation between the coupling constant $L$ and the standard Navarro--Frenk--White scale parameter $r_s$, hinting at a compelling link between these two lengths.
\end{abstract}

\maketitle

\pagebreak
\section{Introduction}
\label{sec: Intro}

A notable topic of interest in modern cosmology is to understand Dark Matter (DM) and Dark Energy (DE), whose origin and nature remain elusive. These components hold a significant importance in our Universe, making up a substantial 95\% of its energy-matter composition ($\sim 68\%$ as DE and $\sim 27\%$ as DM) \cite{Planck:2018vyg}.  
The existence of DM was first inferred in the 1930s by Zwicky. He noticed a discrepancy between the observed dynamical mass of galaxy clusters and the mass derived from theoretical calculations \cite{Zwicky:1937zza}. Subsequent pioneering studies on the rotation curves of spiral galaxies confirmed this inconsistency at galactic scales \cite{Rubin:1970zza}. These results challenged the previously held assumption that the concentration of a galaxy's mass is within its central bulge, which contains the majority of its stars and gas. Instead, observations indicated a notably more consistent density extending across the entire gravitational structures. This ``hidden'' mass which could not be directly observed thus acquired the name of ``dark matter''. 

Since then, more and more evidence has accumulated supporting the need of DM on both cosmological and astrophysical scales. The enthusiastic exploration driven by particle physics considerations has generated numerous models and a variety of dark matter candidates (check \cite{Bertone:2004pz, Feng:2010gw} for more information). However, up to the present moment, there has been no evidence supporting the existence of these proposed dark matter particles, and the sole means to detect the presence and characteristics of dark matter comes from observations within the realm of astrophysics and cosmology.

Currently, the dominant cosmological model is the $\Lambda$--Cold Dark Matter model ($\Lambda$CDM), which is fully based on General Relativity (GR). While the $\Lambda$CDM  model has been successful in explaining so many observations and closely matches the data we have, it suffers from some problems \cite{Akrami:2018vks, eBOSS:2020yzd, DES:2021esc, DES:2021es, Bull:2015stt, Martin:2012bt, Perivolaropoulos:2021jda, DiValentino:2021izs}.
In the pursuit of understanding the nature of DM and DE, researchers have proposed exploring extended theories of gravity (ETGs) \cite{Clifton:2011jh, Koyama:2015vza}. By embracing the idea that GR is a special case of a more comprehensive theory, we naturally arrive to the realm of ETGs. In certain ETG scenarios, both geometry and matter can undergo modifications, offering us a wider horizon through which to comprehend DM and DE. Over time, a multitude of models have been proposed within the ETG category, each contributing uniquely to our understanding of these phenomena \cite{Li:2011sd, Ishak:2018his, Capozziello:2011et}. 

In this paper, our objective is to investigate the characteristics of the DM fluid considering it to be non-minimally coupled with gravity. One might inquire about the rationale for choosing our perfect fluid to be Non-Minimally Coupled (NMC). In the fundamental concept of fluids in GR, we assume that when we transition from individual particles to a fluid, we are dealing with very small scales that can be expressed with a good approximation as a flat spacetime. However, it is worth exploring the scenario where the scale of the fluid's mean free path is comparable to the scale at which spacetime curvature undergoes changes. This scenario is quite likely to be applicable to DM, which doesn't interact with anything, exhibiting at most only very week (self)interactions. As a result, its mean free path can be potentially as large as the Hubble scale ($l_{\mathrm{{mfp}}} \sim 10^3\, \mathrm{Gpc}$). 
As a consequence, DM can be NMC, and, consequently, the standard Einstein equations will be modified \cite{Bettoni:2015wla, Bettoni:2011fs, Bettoni:2012}. 

Within the framework of GR, there exists an option to contemplate DM as a Bose-Einstein condensate (BEC) \cite{Bettoni:2013zma}; this approach provides a way to satisfy the aforementioned criteria, as a BEC naturally possesses a characteristic length scale\footnote{To know more about the BEC applications to DM the interested reader may check  \cite{Ji:1994xh, Boehmer:2007um, Fukuyama:2007sx, Harko:2011xw, Harko:2011jy, Chavanis:2011uv, Chavanis:2011zi, Rindler-Daller:2012mlr}. It is important to mention that these papers examined the dynamics of a non-relativistic condensate, which differs from our current assumptions.}. Additionally, within the domain of ETGs, modifications to gravity inspired by Born-Infeld theory result in the same modifications \cite{BeltranJimenez:2017doy}.

Furthermore, it has been shown that models assuming  NMC between DM and gravity can resolve the core-cusp problem of $\Lambda$CDM model \cite{Harko:2011xw}. This controversy arises from the contrasting findings between cosmological simulations and observations. Cosmological simulations predict that the density distribution of DM near the center of galaxies will exhibit a cusp-like pattern.  In contrast, observations on dwarf galaxies have revealed different results, demonstrating a  linear increase in velocity as one moves toward their centers, resulting in the eventual development of a central density core (see \cite{NFW:1996, Lokas:2000mu, Boylan-Kolchin:2003xvl, deblock:2010i}).

In the Newtonian limit, NMC manifests as an adjustment to the Poisson equation. 
Specifically, it introduces an additional term proportional to the dark matter density $ \rho $, expressed as $ L^2 \nabla^2 \rho $, thereby the modified Poisson equation now relies not only on density but also on the gradients of the density \cite{Bettoni:2013zma}.

While corrections to Poisson's equation have been examined on stellar scales \cite{Casanellas:2011kf, Pani:2012qb}, there is currently a gap in the analysis at galactic and cluster scales. Here, we try to address this gap by conducting an investigation of these corrections specifically at the scale of galaxy clusters.
We will investigate the implications of our modified Poisson equation without making any initial assumptions about the source or magnitude of $L$.

The structure of our paper is as follows. First, in Sec.~\ref{sec: theory}, we provide a concise overview of the NMC DM model's underlying theory. Following that, we introduce the fundamental principles of gravitational lensing theory, upon which we base our theoretical predictions. Lastly in this section, we shortly review the specifics of our chosen mass density profile. In Sec.~\ref{sec: Data} we go through the data set from the CLASH program that has been utilized in our analysis. In Sec.~\ref{sec: Stat} we outline the key aspects of the statistical analysis we have conducted. Finally, in Section~\ref{sec: Dis}, we provide a comprehensive discussion of our results and present our concluding findings.

\section{Theory}
\label{sec: theory}

We will quickly go over our model's theoretical foundation in this section (see \cite{Bettoni:2015wla,Gandolfi:2021jai} for more detail).

The general action that describes the NMC case can be written as follows \cite{Bettoni:2015wla} 
\begin{equation}
\label{eq: action}
    \begin{split}
        S = &\frac{M_{\textrm{Pl}}^2}{2}\int d^4 x\sqrt{-g}\big[R+\alpha_{\mathrm{c}} \rho_c(n,s)R \\
        &+\alpha_{\mathrm{d}} \rho_d(n,s) R_{\mu\nu}u^\mu u^\nu\big] +S_{\mathrm{fluid}}\,,
    \end{split}
\end{equation}
where, the action $S_{\mathrm{fluid}}$  represents the behavior of dark matter which we will model using the action for a perfect fluid. This reads \cite{Bettoni:2015wla, Brown:1992kc}
\begin{equation}
    \label{eq: actionFluid}
    S_{\mathrm{fluid}} = \int d^{4}x \, \sqrt{-g} \rho(n,s) + J^{\mu}(\psi_{,\mu} + s \theta_{,\mu}+ \beta_{A}\alpha^{A}_{,\mu}) ,
\end{equation}
where $n$ represents the particle number density and $s$ indicates the entropy assigned to each particle. In addition, $\alpha^{A}$ and $\beta_{A}$, where $A$ takes on values of 1, 2, 3, represent the Lagrangian coordinates for the fluid. The second term introduces some limitations on the perfect fluid's flow. In addition, $\psi$ and $\theta$, have a thermodynamic interpretation in terms of thermodynamic potentials. Furthermore, $J^{\mu}$ is defined as
\begin{equation}
\label{eq: J}
    J^{\mu} = n u^{\mu} \sqrt{-g}\,,
\end{equation}
and is the conserved current representing particles number conservation, being
$u^{u}$ the four-vector velocity of the fluid. 

In Eq.\eqref{eq: action}, the term $\rho_c(n,s)R$ represents a conformal coupling term, while $\rho_d(n,s)R_{\mu\nu}u^\mu u^\nu$ shows a disformal one where our fluid variable couple to the contracted Ricci tensor with the fluid four-vector velocity. 

As we will explain below, we are focusing specifically on utilizing the disformal coupling term only. Hence, we retain only the latter term in the total action
\begin{equation}
\label{eq: actiondis}
    S=\int d^4x\sqrt{-g}\left[\frac{M_{\textrm{Pl}}}{2}\left(R+\alpha_{d} \rho_d(n,s)R_{\mu\nu}u^\mu u^\nu \right)\right]+S_{\mathrm{fluid}}\,.
\end{equation}

In order to derive the Newtonian limit of our theory, it is beneficial to employ fluid approximation \cite{Bettoni:2012}. This approach leads to a modification of the Poisson equation of the form \cite{Bettoni:2015wla, Bettoni:2011fs}
\begin{equation}
\label{eq: poissonmod}
    \mathbf{\nabla}^2\Phi =4\pi G_N\,[(\rho+\rho_{\rm bar}) - \epsilon\,L^2\,\nabla^2\rho]~,
\end{equation}
where $\Phi$ represents the Newtonian potential, and $\rho_{\rm bar}$ and $\rho$ denote the mass densities of baryonic matter and dark matter, respectively. The second term of the Eq.\eqref{eq: poissonmod} represents the non-minimal coupling term, where $L$ denotes the non-minimal coupling length and $\epsilon= \pm 1$ represents the polarity of the coupling. According to \cite{Gandolfi:2021jai}, the negative polarity $\epsilon=-1$ is required.
In our chosen scenario for disformally coupled fluid, we only have one gravitational potential and no anisotropic stress. Additionally, it's important to highlight that Eq.\eqref{eq: poissonmod} is not the most general expression and having extra terms is possible, but most of them have to be close to zero to satisfy the equivalence principle \cite{DiCasola:2013iia, Bekenstein:1992pj}.
 
It is worth mentioning that this kind of modification to the dynamics of GR (Eq. \ref{eq: actionFluid}) is linked to a ``coarse-grained" scenario and doesn't involve making fundamental modifications to gravitational dynamics. Consequently, $L$ is not a fundamental constant of nature and its value can hence depend on local environment.

As can be seen in Eq.\eqref{eq: poissonmod}, the modified Poisson equation has a term that is dependent on gradients of density. As a consequence, the impact of this modification becomes more pronounced as the distribution of dark matter becomes increasingly inhomogeneous \cite{Bettoni:2011fs}.

According to \cite{Gandolfi:2022puw}, this modification has an important role in modifying the dynamics of spiral galaxies. In addition, in the mentioned study it has been shown that NMC DM can provide a better fit to the rotation curves of spiral galaxies than NFW.

\subsection{Gravitational Lensing}

Gravitational lensing emerges as a potent tool for probing the distribution of both dark and baryonic matter within galaxy clusters. 

Considering a source that is positioned at an angular diameter distance of $D_A$, $D_s$ is the distance from the observer and $D_{l}$ would be distance from the lens; the distance between the lens and the source is denoted as $D_{ls}$ in a gravitational lensing setup \cite{Meneghetti2021LNPbook,Bartelmann:1999yn, Umetsu:2020wlf}.
The angular diameter distance which is a function of redshift can be defined as 
\begin{equation}
\label{eq: distance}
D_A(z) = \frac{c}{1+z}\int_0^z \frac{dz'}{H(z')}\, ,
\end{equation}
where, in the context of a $\Lambda$CDM (Lambda Cold Dark Matter) model, the Hubble function denoted as $H(z)$ is expressed through the first Freedman equation:
\begin{equation}
    H(z) = H_{0} \sqrt{\Omega_{m} (1+z)^{3} + \Omega_{k} (1+z)^{2} + \Omega_{\Lambda}}\, .
    \nonumber
\end{equation} 
with $\Omega_{\Lambda} = 1-\Omega_m$ in the case of spatial flatness ($\Omega_k =0$). Throughout this work, we are assuming our background cosmology parameters to be given from \textit{Planck} baseline model \cite{Planck:2018vyg}, with the values: Hubble constant $H_0 = 67.89$ km s$^{-1}$ Mpc$^{-1}$ and the matter density parameter $\Omega_m = 0.308$. Additionally, we assume that this system can be roughly thought of as two-dimensional given the scale differences between $D_l$ and $D_{ls}$ distances compared to the physical dimensions of the lens (the ``thin-lens" approximation)\footnote{The lens equation is as follows:
\begin{equation}
    \label{eq:lens_eq}
    \vec{\beta} = \vec{\theta} - \frac{D_{ls}}{D_s}\, \hat{\vec{\alpha}}(\vec{\theta})\,,
    \nonumber
\end{equation}
where $(\vec{\beta})$ is the angular position of the source and $(\vec{\theta})$ is the angular position of the observer.}.
In such case, the lens's primary function is to deflect light beams from the source by an angle called $\hat{\vec{\alpha}}$, which is defined as
\begin{equation}
   \hat{\vec{\alpha}}= \frac{2}{c^{2}} \int_{-\infty}^{+\infty} \vec{\nabla}_{\perp} \Phi \mathrm{d}z\, ,
\end{equation}
where $\vec{\nabla}_{\perp}$ represents the two-dimensional gradient operator, which is perpendicular to the path of the light. Additionally, $z$ denotes the coordinate that specifies the position along the path in which the light is propagating.

The deflection angle $\hat{\vec{\alpha}}$, can be described using the effective lensing potential
\begin{equation}
    \label{eq: poten}
    \Phi_\mathrm{lens}(R) = \frac{2}{c^2}\frac{D_{ls}}{D_lD_s}\int^{+\infty}_{-\infty}\Phi(R,z)dz \, ,  
\end{equation}
where $R$ is the two-dimensional projected radius on the lens plane.

The Laplacian of Eq.(\ref{eq: poten}) gives twice the lensing convergence
\begin{equation}
    \kappa(R) = \frac{1}{c^2}\frac{D_{ls}D_l}{D_s}\int^{+\infty}_{-\infty}\Delta_r\Phi(R,z)dz \, ,
\end{equation}
where as mentioned earlier $R$ is the two-dimensional projected radius in the lens plane, $r = \sqrt{R^2+z^2}$ is the three-dimensional radius and $\Delta_r = \frac{2}{r}\frac{\pa}{\pa r} + \frac{\pa^2}{\pa^2 r}$ represents the radial Laplacian in spherical coordinates, where we assume spherical symmetry for simplification. Now, by using the standard Poisson equation
\begin{equation}
    \Delta_r\Phi = 4\pi G_N\rho(r),
\end{equation}
we can establish a connection between the convergence $\kappa$ and the distribution of mass density $\rho$ within the lens system, ultimately leading us to another expression for convergence
\begin{equation}
    \label{eq: kappa}
    \kappa (R) = \int_{-\infty}^{+\infty} 
    \frac{4 \pi G_N}{c^{2}} \frac{D_{ls}D_{l}}{D_{s}}
    \rho(R,z)\mathrm{d}z \equiv \frac{\Sigma(R)}{\Sigma_{cr}}\, ,
\end{equation}
where $\Sigma(R)$ is the lens's two-dimensional surface density and $\Sigma_{cr}$ is the critical surface density of gravitational lensing and these quantities are expressed as follows, respectively
\begin{equation}
    \Sigma(R) = \int_{-\infty}^{+\infty} \rho(R,z) \mathrm{d}z\, ,
\end{equation}
\begin{equation}
   \Sigma_{cr} \,= \frac{c^{2}}{4\pi G_N} \, \frac{D_{s}}{D_{ls}D_{l}}\, .
\end{equation}

It should be mentioned that, till this point, our focus has been on the GR case, thus having $\Phi = \Psi$. Yet, we can broaden our perspective by considering a more general scenario where we have non-zero anisotropic stress, which means the gravitational ($\Phi$) and metric potential ($\Psi$) are not equal ($\Phi \neq \Psi$). In this context the expression for the convergence can be generalized to
\begin{equation}
 \label{eq: kappamod_1}
 \kappa(R) = \frac{1}{c^2}\frac{D_{ls}D_l}{D_s}\int^{+\infty}_{-\infty}\Delta_r\biggl\{\frac{\Phi(R,z) + \Psi(R,z)}{2}\biggr\}dz\, .
\end{equation}

For the model we are focusing on and outlined in Eq. \eqref{eq: poissonmod}, in the case of a disformal coupling, we do have $\Phi = \Psi$ but with the modified Poisson equation Eq.~(\ref{eq: poissonmod}), for which we will have the following convergence for our model
\be
\label{eq: kappamod_2}
   \kappa (R) = \frac{1}{\Sigma_{cr}} \int_{-\infty}^{+\infty} 
    \left[\rho(R,z) - \epsilon L^2 \Delta_r \rho_{NFW}(R,z)\right] \, \mathrm{d}z 
     \, ,
\ee
where $\rho= \rho_{NFW} + \rho_{gas}$ is the total density. In the case of DM, as described in the next section, we considered the Navarro-–Frenk-–White to describe its density profile; for the hot intracluster gas component, we used the profile $\rho_{gas}$ described in the following pages.

As can be seen, the new convergence is influenced by the behavior of $\rho(R,z)$ and the radial Laplacian of only DM, $\Delta_r \rho_{NFW}(R,z)$. Consequently, notable variations in the density profile may impact the value of the convergence. One should also consider the influence of the parameter $L$, distinct for each cluster, which further contributes to the overall change in convergence.

\subsection{Navarro-Frenk-White profile}

The mass distribution within galaxy clusters is frequently represented using spherically symmetric Navarro--Frenk--White (NFW) mass density profile \cite{NFW:1996}. One might argue that such a distribution emerges from simulations in the context of standard General Relativity. We thus follow a minimally-conservative approach, in which we explore if the NFW profile is compatible with the modified scenario and still can be used as density profile for DM distribution in galaxy clusters. But we are aware that the only way to check if a different DM distribution would be achieved in the modified scenario we are considering here, would be to run cosmological simulations based on it. But this is out of the scope of this work.

It's important to note that, in this study, we assume that the mass distribution in galaxy clusters is primarily influenced by DM
\begin{equation}
\label{eq: NFW}
    \rho_\mathrm{NFW}(r) =  \frac{\rho_{s}}{\frac{r}{r_{s}} \Big( 1 + \frac{r}{r_{s}} \Big)^{2}} \,,
\end{equation}
where $\rho_{s}$ represents the characteristic density of the halo, while $r_{s}$ corresponds to the scale radius. Moreover, $\rho_{s}$ can be written
\begin{equation}
\label{eq: delta}
    \rho_{s} = \frac{\Delta}{3} \rho_{c} \frac{c_{\Delta}^{3}}{\mathrm{ln}(1+ c_{\Delta}) - \frac{c_{\Delta}}{1 + c_{\Delta}}} \, ,
\end{equation}
where 
\begin{equation}
 c_{\Delta} = \frac{r_{\Delta}}{r_{s}}\, ,   
\end{equation}
$c_{\Delta}$ --the ratio of the size of the halo-- is the dimensionless concentration parameter. $r_{\Delta}$ represents the spherical radius where the average density inside it is equal to $\Delta$ times the critical density $\rho_c$ of the Universe at the redshift of the lens which here is the cluster. In addition, we have also $M_{\Delta}$, which corresponds to the total mass encompassed within the overdensity radius $r_{\Delta}$
\begin{equation}
\label{eq: M_delta}
    M_{\Delta} = \frac{4}{3} \pi r_{\Delta}^{3} \Delta \rho_{c} = 4 \pi \rho_{s} r_{s}^{3} \bigg[ \mathrm{ln}(1+ c_{\Delta}) - \frac{c_{\Delta}}{1 + c_{\Delta}} \bigg]\, .
\end{equation}
For our analysis, we have fixed the value of $\Delta$ to be 200. As a result, the free NFW parameters we have utilized in our study are $\{c_{200}, M_{200}\}$.

\subsection{Hot gas}
\label{sec:Gas}

Although it would be possible to use X-ray observations for the CLASH clusters, which all have related archival data \cite{Donahue:2014qda}, we have decided to not take directly into account them. As it is well known, such type of observables might be biased by non-gravitational local astrophysical phenomena, contrarily to gravitational lensing, which is a neat gravitational probe. Thus, we have decided to sacrifice a bit of precision (X-ray reconstructed masses are generally better than some lensing-based data) for a stronger and lesser biased reconstruction.  

Despite this, we consider hot gas in our modelling of the clusters, and we include $\rho_{gas}$ in the total density appearing in Eq.~(\ref{eq: kappamod_2}). From the data at our disposal (as discussed in the next section), we fit the gas densities with a double (truncated) $\beta$ model
\begin{align}\label{eq: gas_dens}
\rho_\mathrm{gas}(r) &=  \rho_{e,0}\biggl(\frac{r}{r_0}\biggr)^{-\alpha}\biggl[1 + \biggl(\frac{r}{r_{e,0}}\biggr)^2\biggr]^{-3\beta_0/2} \nonumber \\
& + \rho_{e,1}\biggl[\biggl(\frac{r}{r_{e,1}}\biggr)^2\biggr]^{-3\beta_1/2}\, . 
\end{align}
Note that the free parameters in this expression are fixed at a preliminary stage, by independent fits, and are not left free in the global analysis.

\section{Data}
\label{sec: Data}

In this study, we have used the data from the CLASH (Cluster Lensing And Supernova survey with Hubble) program\footnote{\url{https://archive.stsci.edu/prepds/clash/}} \cite{CLASH2012}. 

The goal of the survey was (among others) to analyse the gravitational lensing characteristics of a set of massive galaxy clusters selected in the redshift range $0.18<z<0.90$ to precisely determine their mass distributions. The sample covers a wide range of masses,  $5\lesssim M_{200}/10^{14}M_\odot\lesssim 30$, and each cluster has both weak- and strong-lensing data from \textit{Hubble} Space Telescope focusing on the central regions \cite{Merten:2014wna,Zitrin2015} combined with ground-based weak-lensing shear and magnification data from the Subaru Telescope \cite{Umetsu:2014vna}. The radial convergence profiles for 20 clusters \cite{Umetsu:2015baa} is then reconstructed. Out of these 20 clusters, 16 were selected based on X-ray observations, while 4 were chosen through lensing observations.

Our work focuses on a subset of the CLASH sample, consisting of 15 clusters selected based on X-ray observations and 4 clusters chosen through lensing observations, as described in \cite{Umetsu:2015baa}. One of the X-ray-selected clusters, RXJ1532, was excluded from our analysis because its mass reconstruction was based only on wide-field weak-lensing data resulting in too large errors \cite{Zitrin2015}. % 
The clusters in our analysis sample span a redshift range of $0.187 \le z \le 0.686$, with a median redshift of $z_\mathrm{med}= 0.352$. The resolution limit of the mass reconstruction, determined by the \textit{HST} lensing data, is typically around 10 arcseconds ($\approx 35 h^{-1}$ kpc) at the median redshift \cite{Meneghetti:2014xna}. It is worth noting that approximately half of the selected clusters in our sample are anticipated to be unrelaxed \cite{Meneghetti:2014xna}.

In \cite{Umetsu:2015baa}, it is mentioned that the average surface mass density ($\Sigma(R)$) of the X-ray-selected subset from the CLASH sample is most accurately described by the NFW profile when considering GR. The NFW model is effective in explaining the distribution of dark matter in clusters, as it dominates the overall cluster scale. On the other hand, cluster baryons, including X-ray-emitting hot gas and BCGs, are influenced by non-gravitational and local astrophysical phenomena. Consequently, estimates of the total mass based on hydrostatic methods using X-ray observations are heavily influenced by the dynamic and physical conditions within the cluster. In comparison, gravitational lensing offers a direct means to investigate the projected mass distribution in galaxy clusters.

\section{Statistical analysis}
\label{sec: Stat}
In order to constrain the values in the non-minimally coupled model and the variables describing the NFW profile for each cluster, we need to define a $\chi^2$ function. Therefore, $\boldsymbol{\theta} = \{c_{200}, \, M_{200}, \, L \}$ denotes the collection of parameters we see as variables in our theory. Surely, as we switch into the realm of GR, this will change to $\boldsymbol{\theta} = \{c_{200}, \, M_{200} \}$. The $\chi^2$ function is defined as below 
\begin{equation}
\label{eq: chi}
    \chi^{2} = \big( \boldsymbol{\kappa^{theo}}(\boldsymbol{\theta}) - \boldsymbol{\kappa^{obs}} \big) \cdot \mathbf{C}^{-1} \cdot \big( \boldsymbol{\kappa^{theo}}(\boldsymbol{\theta}) - \boldsymbol{\kappa^{obs}} \big)\, ,
\end{equation}
where $\boldsymbol{\kappa^{obs}}$ refers to the data vector related to the observed convergence values. This vector comprises 15 data elements, each corresponding to the measured value of $\kappa$ in a specific radial bin. The vector $\boldsymbol{\kappa^{theo}}(\boldsymbol{\theta})$ contains the theoretical predictions for the convergence of the model, calculated using Eq.~(\ref{eq: kappa}). Additionally, $\mathbf{C}$ represents the covariance error matrix \cite{Umetsu:2015baa,Umetsu:2020wlf}.

We employed our custom Monte Carlo Markov Chain (MCMC) code to minimize the $\chi^{2}$ function. To ensure the convergence of the chains, we followed the approach described in \cite{Dunkley:2004sv}. 
To assess the credibility of our NMC model compared to standard GR by a meaningful statistical comparison, we calculated the Bayesian Evidence \cite{InformationTheory}, $\mathcal{E}$, for both models for each of cluster  using the nested sampling algorithm explained in \cite{Mukherjee:2005wg}. Since the selection of priors may significantly impacts Bayesian evidence \cite{Nesseris:2012cq}, we maintained consistency by always choosing the same uninformative flat priors for the parameters.

The posterior distribution $\mathcal{P}(\boldsymbol{\theta},\mathcal{M}|{D})$, which we get as output from our MCMCs, is defined as
\begin{equation}
\label{eq: evidence_1}
   \mathcal{P}(\boldsymbol{\theta},\mathcal{M}|{D}) = \frac{\mathcal{L}(D|\boldsymbol{\theta},\mathcal{M}) \pi(\boldsymbol{\theta},\mathcal{M})}{\mathcal{E}(D|\mathcal{M})},
\end{equation}
where $\boldsymbol{\theta}$ is the set of parameters of our models $\mathcal{M}$ (GR and NMC), having the data $D$, and $\mathcal{L}(D|\boldsymbol{\theta},\mathcal{M}) \propto \exp (-\chi^2 (\boldsymbol{\theta})/2$) is the likelihood distribution function given the priors distributions $\pi(\boldsymbol{\theta},\mathcal{M})$.
Thus, the Evidence is
\begin{equation}
\label{eq: evidence_2}
   \mathcal{E}(D|\mathcal{M})= \int d\boldsymbol{\theta}\mathcal{L}(D|\boldsymbol{\theta},\mathcal{M})P(\boldsymbol{\theta}|\mathcal{M})\, .
\end{equation}
We calculate the Bayes Factor ($\mathcal{B}^{\,i}_{j}$), defined as the ratio of evidence values between two models
\begin{equation}
\label{evidence_1}
    \mathcal{B}^{\,i}_{j} = \frac{\mathcal{E}(M_{i})}{\mathcal{E}(M_{j})}\;,
\end{equation}
with $\mathcal{M}_{j}$ being the reference model (in our case, GR). The comparison of models is then conducted employing the empirically calibrated Jeffreys scale \cite{JeffreysScale} which states that: if $\ln\mathcal{B}_{ij} < 1$, the evidence in favor of model $i$ is weak against model $j$; if $1 < \ln\mathcal{B}_{ij} < 2.5$ the evidence is substantial; if $2.5 < \ln\mathcal{B}_{ij} < 5$ it is strong; if $\ln\mathcal{B}_{ij} > 5$ it becomes decisive.

Moreover, in order to be sure to minimize the impact from the applied priors on the Bayesian comparison, we have also resorted on the Suspiciousness, $ \mathcal{S}^{i}_{j}$, introduced in \cite{Handley:2019wlz,Handley:2019pqx,Joachimi:2021ffv} and defined as
\be
\label{eqn: sus}
\ln\mathcal{S}^i_j = \ln\mathcal{B}^i_j + \mathcal{D}_{KL,i} - \mathcal{D}_{KL,j}
\ee
where $\mathcal{D}_{KL}$ is the Kullback-Leibler (KL) divergence \cite{10.1214/aoms/1177729694}. An interpretation of the suspiciousness, similar to Jeffrey's scale for the Bayes Ratio, is provided by Fig. 4 of \cite{Joachimi:2021ffv}. Specifically, a negative value of $\log\mathcal{S}^i_j$ should be intended as a sign of tension; a positive value of $\log\mathcal{S}^i_j$ instead as a sign of concordance.

We anticipate here (a discussion of the reasons behind our choice will be detailed in the next section) that in our statistical analysis we have chosen two different approaches: ``standard'' marginalisation, as working directly on the MCMCs outputs; and the profile distribution (PD - an extension of the profile likelihood) \cite{Gomez-Valent:2022hkb, Trotta:2017wnx} approach. 

\section{Results and Discussion}
\label{sec: Dis}

In Table~\ref{tab: results_1} we report all the main results of our analysis. In the first step, in the second and third column we show the values for the NFW parameters $c_{200}$ and $M_{200}$ in the GR case, which will serve as our benchmark model, and we find a perfect agreement (just as cross-check of our codes) with results from literature \cite{Bouche:2022jts, Umetsu:2015baa}. In Table~\ref{tab: results_2} we report some secondary (not directly fitted) quantities which are equally important: the characteristic lengths from GR, namely, $r_{200}$ and the NFW scaling, $r_s$.

In both tables, regarding our NMC model, the results we report are obtained both from custom marginalisation, namely, simply ``reading'' the posteriors which are produced as output by the MCMCs; and after applying a PD procedure. The reason for this double analysis is due to clear volume-effects which we can notice when we take a more close inspection of the $\chi^2$ (or, equivalently, $\mathcal{L}$) landscape. Indeed, one can easily spot that the results obtained by standard marginalisation for our NMC model exhibit just a statistically not-significant deviation from GR for what concerns the NFW parameters, $c_{200}$ and $M_{200}$. On the other hand, when looking carefully at the posterior distribution of the main characterizing NMC parameter, the coupling length $L$, we note how its peak is generally highly shifted from the value at which we effectively get the minimum $\chi^2$, which does should serve as best fit estimation for this parameter. Actually, the region around the minimum is poorly explored with respect to rest of the parameter space because it is quite narrow, thus, volume-effects might be penalizing the physical information we may infer from our analysis and jeopardize our final assessments about the reliability of the NMC model with respect to GR. The PD approach as described in \cite{Gomez-Valent:2022hkb} is designed exactly to highlight statistical inference beyond such volume-effects.  

Before any conclusion can be drawn, it is important to highlight that the value of $r_{200}$, as it is possible to check from Table~\ref{tab: results_2}, does not change in a statistically significant way when moving from GR to the NMC case. That is important, because it means that the scale at which the concentration and the mass are estimated are the same in both cases and, thus, any difference can be consistently compared.

\begin{figure*}[t!]
\centering
\includegraphics[width=8.5cm]{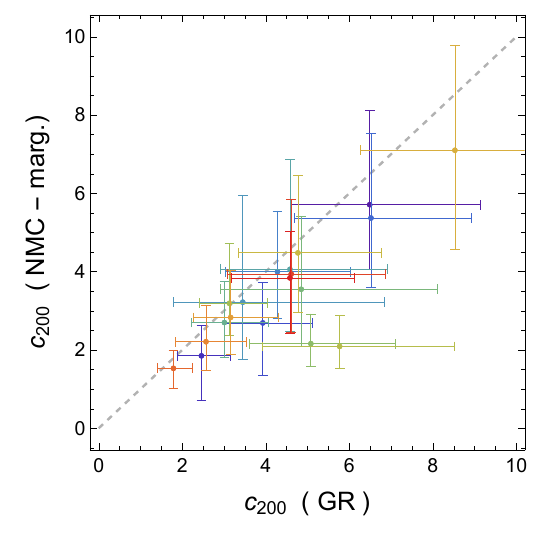}~~~
\includegraphics[width=8.5cm]{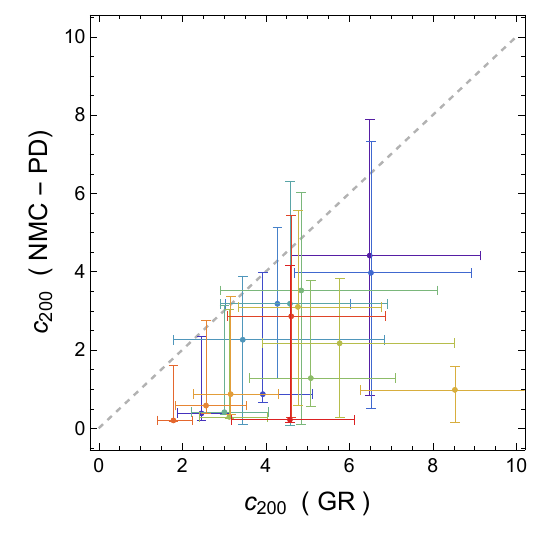}\\
\includegraphics[width=8.5cm]{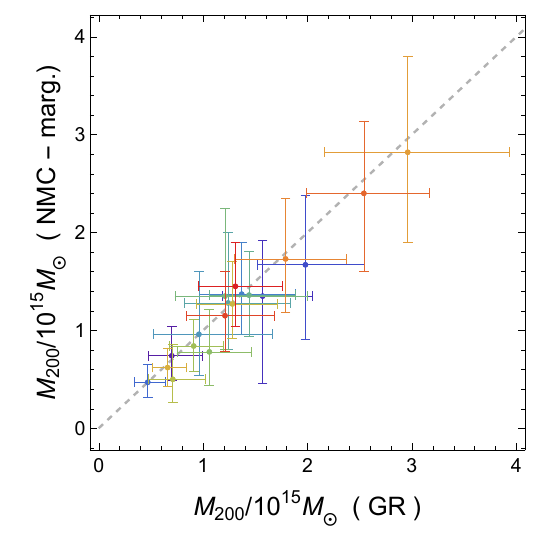}~~~
\includegraphics[width=8.5cm]{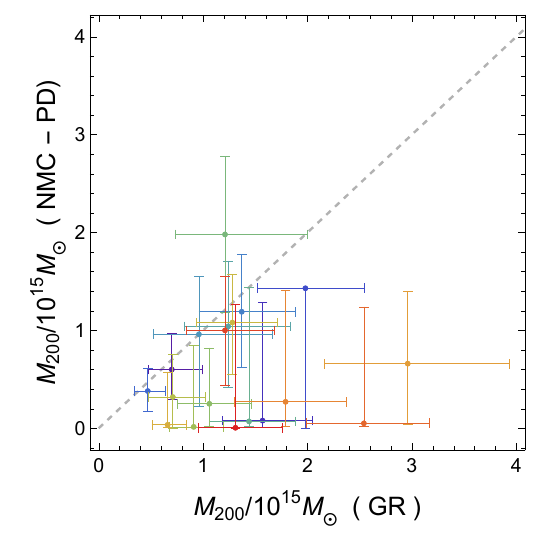}
\caption{Comparison of the constraints on dark matter parameters for $c_{200}$ and for $M_{200}$ obtained from GR and from the NMC model considered in this work. In the left panels, we plot results from the marginalisation procedure; in the right ones, we show results from the profile distribution procedure.}
\label{fig: compare_GR}
\end{figure*}

The difference between the marginalisation and the PD approach is made clear in our figures. For example, in Fig.~\ref{fig: compare_GR}, we present a comparison between the values of $c_{200}$ and $M_{200}$ acquired from GR and our NMC model.
The left panels illustrates the comparison in the marginalisation case, while the right panel showcases the PD outcomes.

\begin{figure*}[t!]
\centering
\includegraphics[width=10.cm]{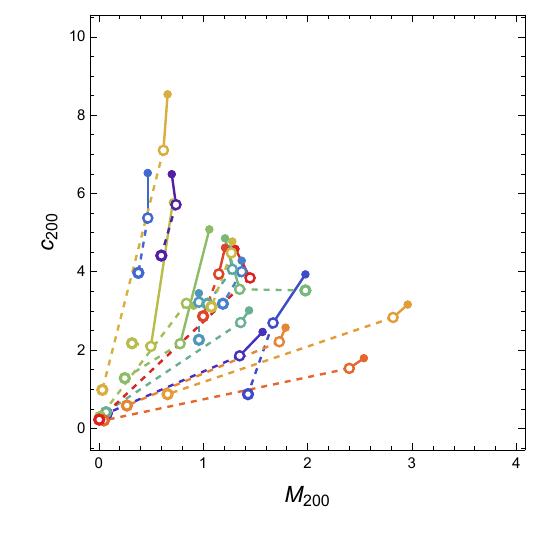}
\caption{Comparison of the constraints on dark matter parameters $\{c_{200},M_{200}\}$ obtained from GR and from the NMC model considered in this work. Solid circles are GR; empty circles are from NMC after marginalisation; bold empty circles are from NMC after profile distribution procedure. Solid lines connect GR and NMC after marginalisation; dashed lines connect NMC after marginalisation with NMC after profile distribution procedure. We avoid to plot error bars for the sake of clarity.}
\label{fig: comparing_GR_NMC}
\end{figure*}

Upon closer look at these figures, a notable trend emerges. In general, the values obtained for $c_{200}$ and $M_{200}$ from the marginal analysis align more closely with those derived from GR, while the PD results exhibit more pronounced variations from the GR predictions. More specifically, in the PD analysis we see how both the concentration and the mass of the NFW profile are systematically lower than the GR case. We try to stress even more this trend in Fig.~\ref{fig: comparing_GR_NMC}, where we do not show error bars for the sake of clarity, and we connect, for each cluster, GR (solid circles) to NMC marginalisation (empty circles) results with solid lines, and NMC marginalisation results to NMC PD ones (bold empty circles) with dashed ones. Thus, the NMC model, at least within the internal $r_{200}$ region, requires less massive and less concentrated dark matter haloes in order to explain lensing data.

It is now interesting to give a look at the parameter which actually characterizes the NMC model, the interaction length $L$ (for numerical reasons, we have chosen to work with $\log{L}$). In the case of a marginalized analysis, the estimated value of $L$ appears to be ``relatively'' small, where the qualitative ``relatively'' should be clarified. Indeed, we are dealing with clusters which have ranges of the order of few Mpc, and $L$ ranges from $0.1$ to $10^2$ kpc, with a typical average value $\sim10$ kpc. The main consequence which could be draw from this result, is that the correction to the Poisson equation introduced by the NMC model is just a small ``perturbation'' to the standard one.    
This result prompts further investigation. In our quest for validation, we can compare our results with those of \cite{Gandolfi:2023hwx}, which addresses similar research objectives, albeit with a different data set. Interestingly, their computed value for $L$ is also very small and of the same order of our finding, as shown in their corresponding Fig.~4.

When moving to the PD analysis, things change substantially. In some case PDs are in full disagreement with the marginalized results: for example, in the case of MACS0416 we move from $L \sim 10^{-2}$ kpc in the marginalized case to $L \sim 1$ Mpc in the PD one. This is a general trend: from the PD analysis, which once again we remind highlights the behaviour of the posterior around the maximum of the likelihood, we get systematically larger values for $L$ with respect to the marginalized analysis. 

\begin{figure*}[t!]
\centering
\includegraphics[width=8.5cm]{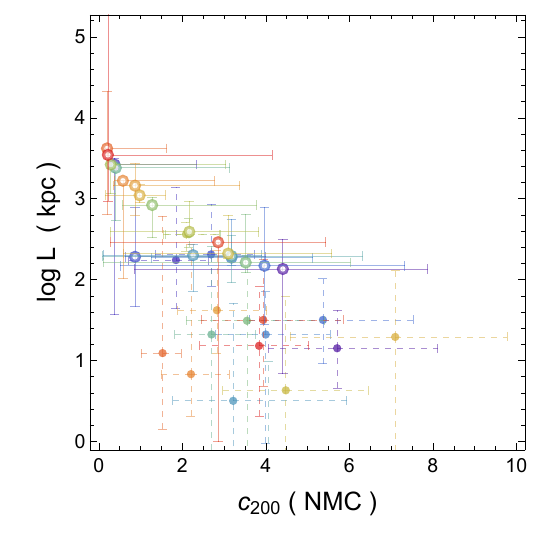}~~~
\includegraphics[width=8.5cm]{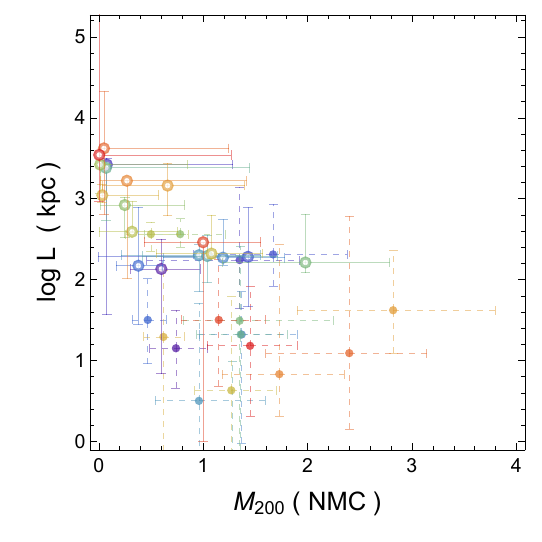}\\
\includegraphics[width=8.5cm]{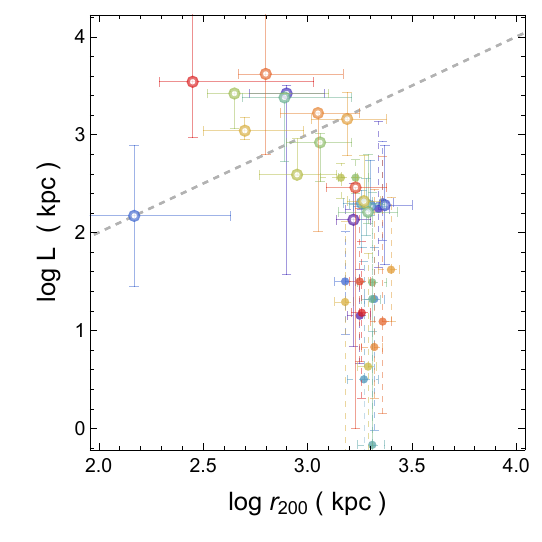}~~~
\includegraphics[width=8.5cm]{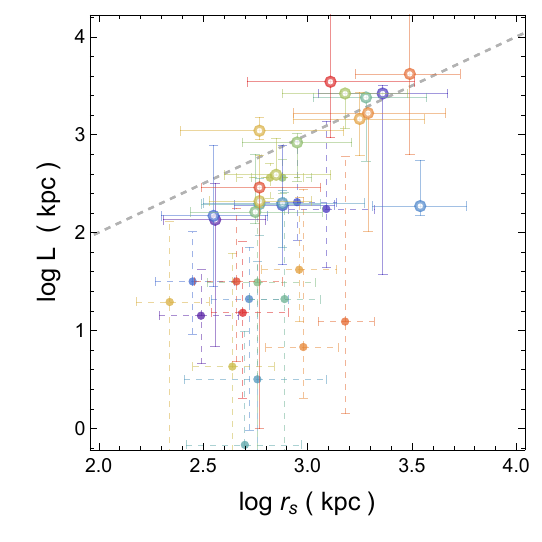}\\
\caption{Comparisons between the NMC characteristic parameter $L$, and the NFW parameters. In the plot, solid circles represent the Marginalized analysis, and empty circles show the PD ones.}
\label{fig: compare_cMr_vs_L}
\end{figure*}

To get even more insight we compare $L$ with the NFW parameters, $c_{200}$ and $M_{200}$ and with the other two characteristic lengths from GR, $r_{200}$ and $r_s$ in Fig.~(\ref{fig: compare_cMr_vs_L}), where the results from marginalized analysis are shown as solid circles and those from PD are empty ones.

In the top left plot, we see again the shift towards smaller values of the concentration which we obtain in the PD analysis, but more strikingly we see an almost perfect anti-correlation between $c_{200}$ and $L$. Although, this is not surprising, because in the corrective term induced by the NMC model into the Poisson equation, Eq.~({\ref{eq: poissonmod}}), we actually have the combination $L^2\,\rho_s$, with $\rho_s$ being mostly dependent on $c_{200}$, as in 
Eq.~(\ref{eq: delta}). The same anti-correlation vs the mass is visible also in the top right panel, although much weaker. Also, we notice how it goes in the opposite direction with respect to what shown in Fig.~4 by \cite{Gandolfi:2023hwx}, although in this reference they do not seem to have performed the PD analysis. 

In the bottom panels of Fig.~\ref{fig: compare_cMr_vs_L} we have the most interesting finding. On the left, we compare $L$ with $r_{200}$, and we notice how moving from the marginalized to the PD analysis leads from $L \ll r_{200}$ to $L \sim r_{200}$. This same pattern is even more clearly evident on the right, where the correspondence between $L$ and $r_s$ seems to be almost perfect, with a tentative weighted fit producing $\log L \approx (0.964 \pm 0.034) \log r_{s} $. If confirmed by further investigations, these results would state for the NFW scaling $r_{s}$ a sort of ``natural'' explanation if connected to the interaction lenght $L$ of dark matter explained as an NMC fluid.

While the anti-correlation between $c_{200}$ and $L$ can be easily explained, this latter correlation is more tricky, and interesting. If we expressed the NFW profile and the NMC correction in dimensionless units, $x=r/r_s$, we would have
\begin{equation}
\rho_{NFW}(x) \propto \frac{1}{x \left(1+x\right)^2} + \frac{L^2}{r^{2}_s} \frac{6}{x\left(1+x\right)^4}\, .
\end{equation}
This would be also expected from dimensional considerations and by the second derivative nature of the NMC. But in no way it \textit{implies} the linear correlation between $L$ and $r_{s}$. If the NMC would contribute in a negligible way, it would be more logical and statistically favoured to expect small values for $L$. Moreover, the correction is itself function of the scale, $r$.

\begin{figure*}[t!]
\centering
\includegraphics[width=8.5cm]{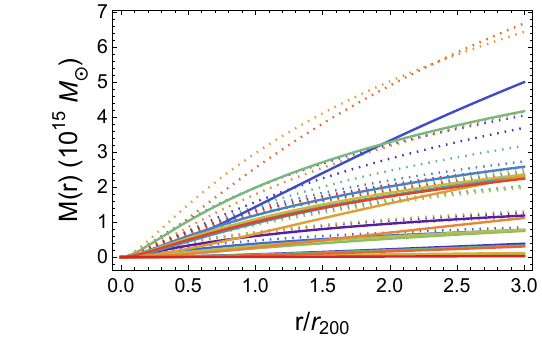}
~~~
\includegraphics[width=8.5cm]{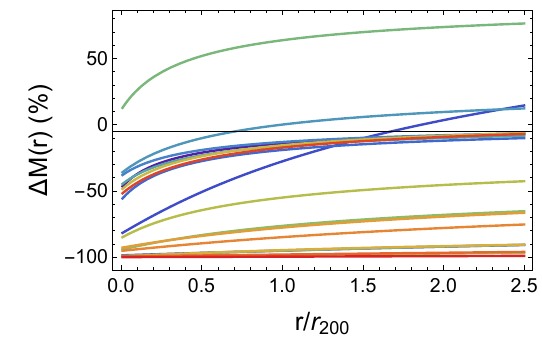}
\caption{Difference in the mass distribution between GR and NMC model from the PD statistical analysis. We plot $M(r)$ (left panel; dotted lines for GR, solid lines for the NMC model) and $\Delta M \equiv (M_{NMC}-M_{GR}) / M_{GR}$ (right panel) as functions of the normalized distance from the center, $r/r_{200}$.}
\label{fig: DeltaM}
\end{figure*}

Finally, in Fig.~\ref{fig: DeltaM} we show the variation in the $M(r))$ distribution from GR to the NMC model at the best fit derived from the PD statistical analysis. Note that $\Delta M_{200}$ is defined as $(M_{200,NMC}-M_{200,GR})/M_{200,GR}$, as considering the change in the scale which is due to the differences in the estimated lenghts, we normalize the distances from the center to $r/r_{200}$. It is quite evident to notice how the NMC model requires much less matter and much less concentrated in most of the cases we have considered: in some cases even more than $70\%$ less dark matter with respect to GR from the inner to the outer regions; in many cases we require half of the mass in the inner regions, with a difference which is less evident $(\approx 10\%)$ at outer ranges; few cases seem to be outliers and deviate from this general trend.

\section{Conclusions}
\label{sec: Conc}

Our investigation centers on exploring the scenario of a non-minimal coupling between dark matter (modeled as a perfect fluid) and gravity. As highlighted earlier, this coupling introduces alterations to the Einstein equations, extending its impact to the Planck mass and the energy-momentum tensor of a fluid, given their reliance on the curvature scale.

By adapting the action and taking the Newtonian limit for the disformal case, one reaches the modified Poisson equation Eq. (\ref{eq: poissonmod}), characterized by an additional term $L^2 \nabla^2\rho$. In this equation, the first term represents the density of dark matter and gas, while the additional term involves the coupling length $L$ and the NFW density $\rho$. 

Leveraging both robust strong and weak gravitational lensing data within the CLASH program, we tested the NMC model across 19 high-mass galaxy clusters. It's noteworthy that our analysis extends beyond dark matter to include the density of gas (X-ray). While the option to incorporate gas data from the CLASH dataset was available, we exercised caution, opting to not consider it due to potential biases.

Our analytical methodology employs two approaches for presenting findings: Marginalisation and Profile Distribution. Recognizing the influence of volume effects in the posterior distribution, we find that the PD is more suitable when working with data.
Applying the PD reveals that dark matter necessitates lower mass and concentration to align with observed lensing data. Furthermore, a noteworthy correlation emerges between the coupling constant $L$ and the standard NFW scale parameter $r_s$ prompting further exploration into the connection between them in future research.

In our forthcoming research, we aim to expand our investigations beyond the exclusive consideration of dark matter and gas. Our focus will encompass additional components, such as galaxies, enhancing our understanding of dark matter characteristics. 

\acknowledgments
The research of S.Z. and V.S. is funded by the Polish National Science Centre grant No. DEC-2021/43/O/ST9/00664. D.B. acknowledge support from  projects PID2021-122938NB-I00 funded by the Spanish “Ministerio de Ciencia e Innovación” and
FEDER “A way of making Europe” and PIC-2022-02 funded by Salamanca University. 

{\renewcommand{\tabcolsep}{1.mm}
{\renewcommand{\arraystretch}{2.}
\begin{table*}[h]
\begin{minipage}{\textwidth}
\centering
\caption{CLASH clusters ordered by redshift. Our results regarding $c_{200}, M_{200}$ and $L$ in GR and Modified version (marginal and PD) for the combination of DM and gas. The units for cluster radii are expressed in kpc.}
\label{tab: results_1}
\resizebox*{\textwidth}{!}{
\begin{tabular}{c|cc|ccc|ccc|cc}
\hline
\hline
 & \multicolumn{2}{c|}{GR} & \multicolumn{3}{c|}{MOD (marg.)} &\multicolumn{3}{c|}{MOD (PD)} & $\log \mathcal{B}^{i}_{j}$ & $\log \mathcal{S}^{i}_{j}$ \\
 Cluster
 & $c_{200}$ & $M_{200}$   
 & $c_{200}$ & $M_{200}$ & $\log L$ 
 & $c_{200}$ & $M_{200}$ & $\log L$ 
 & & \\
 & 
 & $(10^{15}\,\mathrm{M}_{\odot})$ & 
 & $(10^{15}\,\mathrm{M}_{\odot})$ & ($kpc$)  &
 & $(10^{15}\,\mathrm{M}_{\odot})$ & ($kpc$)  
 & & \\
\hline
\hline
A383 &
$6.49^{+2.65}_{-1.89}$ & $0.70^{+0.29}_{-0.22}$ & $5.71^{+2.41}_{-1.65}$ & $0.74^{+0.30}_{-0.25}$ & $1.15^{+0.47}_{-0.49}$ & $4.41^{+3.47}_{-3.56}$ & $0.60^{+0.37}_{-0.30}$ & $(0.84,2.13,2.50)$ &  $0.005^{+0.024}_{-0.023}$ & $-0.010^{+0.035}_{-0.038}$ \\
A209 &
$2.46^{+0.69}_{-0.58}$ & $1.57^{+0.48}_{-0.39}$ & $1.85^{+0.78}_{-1.13}$ & $1.35^{+0.57}_{-0.89}$ & $2.24^{+0.90}_{-0.60}$ & $(0.21,0.38,2.35)$ & $(0.005,0.078,1.285)$ & $(1.57,3.42,3.50)$ &  $0.006^{+0.022}_{-0.022}$ & $-0.025^{+0.038}_{-0.052}$ \\
A2261 &
$3.93^{+1.19}_{-0.92}$ & $1.98^{+0.56}_{-0.46}$ & $2.69^{+1.04}_{-1.33}$ & $1.67^{+0.71}_{-0.76}$ & $2.31^{+0.62}_{-0.39}$ & $(0.65,0.87,3.99)$ & $<1.43$ & $2.28^{+0.61}_{-0.61}$ & $-0.0005^{+0.0190}_{-0.0202}$ & $0.011^{+0.034}_{-0.037}$ \\
RXJ2129 &
$6.52^{+2.41}_{-1.83}$ & $0.47^{+0.17}_{-0.13}$ & $5.37^{+2.17}_{-1.77}$ & $0.47^{+0.18}_{-0.15}$ & $1.50^{+0.51}_{-0.54}$ & $3.97^{+3.36}_{-3.45}$ & $0.38^{+0.23}_{-0.21}$ & $2.17^{+0.72}_{-0.72}$ & $0.014^{+0.022}_{-0.023}$ & $0.005^{+0.040}_{-0.037}$ \\
A611 &
$4.28^{+1.74}_{-1.24}$ & $1.37^{+0.51}_{-0.41}$ & $4.00^{+1.55}_{-1.20}$ & $1.37^{+0.53}_{-0.41}$ & $1.32^{+0.53}_{-1.34}$ & $3.18^{+1.94}_{-1.91}$ & $1.19^{+0.59}_{-0.57}$ & $(2.18,2.27,2.74)$ & $-0.013^{+0.021}_{-0.022}$ & $-0.021^{+0.046}_{-0.041}$ \\
MS2137 &
$3.45^{+3.40}_{-1.67}$ & $0.96^{+0.70}_{-0.44}$ & $3.22^{+2.72}_{-1.45}$ & $0.96^{+0.64}_{-0.42}$ & $0.50^{+1.21}_{-2.49}$ & $(0.11,2.26,3.89)$ & $0.96^{+0.59}_{-0.74}$ & $(1.85,2.30,2.43)$ & $0.057^{+0.021}_{-0.025}$ & $0.084^{+0.031}_{-0.043}$ \\
RXJ2248 &
$4.58^{+2.34}_{-1.67}$ & $1.24^{+0.60}_{-0.42}$ & $4.06^{+2.80}_{-1.58}$ & $1.28^{+0.72}_{-0.47}$ & $-0.17^{+1.16}_{-1.73}$ & $3.18^{+3.13}_{-3.10}$ & $1.04^{+0.67}_{-0.62}$ & $(1.97,2.29,2.55)$ & $-0.049^{+0.017}_{-0.025}$ & $-0.077^{+0.035}_{-0.044}$ \\
MACSJ1115 &
$3.01^{+1.05}_{-0.78}$ & $1.44^{+0.44}_{-0.38}$ & $2.70^{+1.05}_{-0.88}$ & $1.36^{+0.45}_{-0.42}$ & $1.32^{+1.09}_{-3.06}$ & $(0.35,0.41,3.14)$ & $(0.02,0.07,1.44)$ & $(2.73,3.38,3.42)$ & $0.014^{+0.023}_{-0.022}$ & $0.008^{+0.032}_{-0.037}$ \\
MACSJ1931 &
$4.85^{+3.26}_{-1.93}$ & $1.21^{+0.79}_{-0.48}$ & $3.55^{+1.86}_{-1.45}$ & $1.35^{+0.90}_{-0.54}$ & $1.49^{+0.81}_{-2.44}$ & $(0.11,3.52,6.02)$ & $1.98^{+0.80}_{-0.80}$ & $(2.09,2.21,2.80)$ & $-0.025^{+0.027}_{-0.020}$ & $-0.056^{+0.038}_{-0.037}$ \\
MACSJ1720 &
 $5.08^{+2.02}_{-1.48}$ & $1.06^{+0.40}_{-0.31}$ & $2.16^{+0.76}_{-0.57}$ & $0.78^{+0.43}_{-0.34}$ & $2.56^{+0.19}_{-0.16}$ & $(0.57,1.28,3.77)$ & $(0.02,0.25,0.82)$ & $(2.52,2.92,3.01)$ & $0.009^{+0.023}_{-0.022}$ & $-0.016^{+0.046}_{-0.032}$ \\
MACSJ0416 &
$3.13^{+0.90}_{-0.73}$ & $0.91^{+0.28}_{-0.23}$ & $3.19^{+1.54}_{-0.82}$ & $0.84^{+0.27}_{-0.26}$ & $-1.89^{+1.47}_{-3.43}$ & $(0.26,0.29,3.03)$ & $(0.010,0.014,0.850)$ & $(3.06,3.42,3.44)$ & $-0.075^{+0.026}_{-0.022}$ & $-0.143^{+0.046}_{-0.037}$ \\
MACSJ0429 &
$5.77^{+2.75}_{-1.85}$ & $0.71^{+0.31}_{-0.23}$ & $2.09^{+0.80}_{-0.55}$ & $0.50^{+0.36}_{-0.24}$ & $2.56^{+0.15}_{-0.21}$ & $(0.28,2.17,3.82)$ & $(0.002,0.319,0.752)$ & $(2.58,2.59,2.96)$ & $-0.083^{+0.020}_{-0.025}$ & $-0.154^{+0.038}_{-0.046}$ \\
MACSJ1206 &
$4.77^{+2.01}_{-1.43}$ & $1.28^{+0.43}_{-0.34}$ & $4.48^{+1.98}_{-1.51}$ & $1.27^{+0.43}_{-0.35}$ & $0.63^{+1.16}_{-1.23}$ & $3.10^{+2.47}_{-2.51}$ & $1.08^{+0.49}_{-0.53}$ & $(2.26,2.32,2.79)$ & $0.004^{+0.020}_{-0.022}$ & $-0.0004^{+0.0355}_{-0.0402}$  \\
MACSJ0329 &
$8.53^{+2.71}_{-2.26}$ & $0.66^{+0.18}_{-0.15}$ & $7.10^{+2.68}_{-2.52}$ & $0.62^{+0.20}_{-0.19}$ & $1.29^{+0.82}_{-2.30}$ & $0.98^{+0.61}_{-0.83}$ & $(0.006,0.035,0.568)$ & $(2.95,3.04,3.18)$ & $0.009^{+0.020}_{-0.024}$ & $0.0004^{+0.0374}_{-0.0455}$\\
RXJ1347 &
$3.16^{+1.14}_{-0.89}$ & $2.96^{+0.97}_{-0.80}$ & $2.83^{+1.19}_{-0.94}$ & $2.82^{+0.98}_{-0.92}$ & $1.62^{+0.74}_{-0.53}$ & $(0.36,0.87,3.37)$ & $(0.04,0.66,1.40)$ & $(2.79,3.16,3.43)$ & $0.008^{+0.029}_{-0.026}$ & $0.005^{+0.053}_{-0.045}$\\
MACSJ1149 & 
$2.57^{+0.97}_{-0.73}$ & $1.79^{+0.58}_{-0.49}$ & $2.21^{+0.92}_{-0.72}$ & $1.73^{+0.62}_{-0.55}$ & $0.83^{+1.61}_{-0.52}$ & $(0.37,0.58,2.76)$ & $(0.02,0.27,1.41)$ & $(2.01,3.22,3.23)$ & $0.033^{+0.023}_{-0.021}$ & $0.042^{+0.048}_{-0.029}$\\
MACSJ0717 &
$1.79^{+0.46}_{-0.38}$ & $2.54^{+0.63}_{-0.55}$ & $1.53^{+0.46}_{-0.52}$ & $2.40^{+0.74}_{-0.80}$ & $1.09^{+1.69}_{-0.94}$ & $(0.17,0.20,1.61)$ & $(0.02,0.05,1.24)$ & $3.62^{+0.71}_{-0.82}$ & $0.008^{+0.022}_{-0.017}$ & $0.002^{+0.037}_{-0.030}$\\
MACSJ0647 &
$4.61^{+2.26}_{-1.54}$ & $1.21^{+0.47}_{-0.37}$ & $3.94^{+1.91}_{-1.49}$ & $1.15^{+0.45}_{-0.36}$ & $1.50^{+0.75}_{-0.82}$ & $2.86^{+2.57}_{-2.58}$ & $1.00^{+0.55}_{-0.56}$ & $>2.46$ & $0.041^{+0.023}_{-0.020}$ & $0.046^{+0.038}_{-0.034}$\\
MACSJ0744 &
$4.58^{+2.09}_{-1.41}$ & $1.31^{+0.45}_{-0.36}$ & $3.84^{+1.19}_{-1.42}$ & $1.45^{+0.45}_{-0.41}$ & $1.18^{+0.73}_{-0.87}$ & $(0.16,0.22,4.15)$ & $(0.002,0.005,1.267)$ & $(2.97,3.54,3.62)$ & $-0.002^{+0.021}_{-0.023}$ & $-0.003^{+0.032}_{-0.048}$ \\
\hline
\hline
\end{tabular}}
\end{minipage}
\end{table*}}}

{\renewcommand{\tabcolsep}{1.mm}
{\renewcommand{\arraystretch}{2.}
\begin{table*}[h]
\begin{minipage}{\textwidth}
\centering
\caption{CLASH clusters ordered by redshift. Our results regarding $\log r_{200}, \log r_{s}$ and $\log L$ in GR and Modified version (marginal and PD) for the combination of DM and gas. The units for cluster radii are expressed in kpc.}\label{tab: results_2}
\resizebox*{\textwidth}{!}{
\begin{tabular}{c|cc|ccc|ccc}
\hline
\hline
 & \multicolumn{2}{c|}{GR}  & \multicolumn{3}{c|}{MOD (marg.)} & \multicolumn{3}{c}{MOD (PD)} \\
 Cluster 
 & $\log r_{200}$ & $\log r_{s}$   
 & $\log r_{200}$ & $\log r_{s}$ & $\log L$   
 & $\log r_{200}$ & $\log r_{s}$ & $\log L$  \\
 &  & 
 &           &  & 
 &           &  & \\
\hline
\hline
A383 & 
$3.24^{+0.05}_{-0.05}$ & $2.43^{+0.19}_{-0.20}$ & $3.25^{+0.05}_{-0.06}$ & $2.50^{+0.18}_{-0.20}$ & $1.15^{+0.47}_{-0.49}$ & $3.22^{+0.08}_{-0.08}$ & $2.58^{+0.24}_{-0.25}$ & $(0.84,2.13,2.50)$ \\
A209 & 
$3.36^{+0.04}_{-0.04}$ & $2.97^{+0.14}_{-0.14}$ & $3.34^{+0.05}_{-0.16}$ & $3.09^{+0.22}_{-0.18}$ & $2.24^{+0.90}_{-0.60}$ & $2.90^{+0.18}_{-0.18}$ & $3.36^{+0.31}_{-0.31}$ & $(1.57,3.42,3.50)$ \\
A2261 & 
$3.39^{+0.04}_{-0.04}$ & $2.79^{+0.14}_{-0.15}$ & $3.36^{+0.05}_{-0.08}$ & $2.95^{+0.19}_{-0.16}$ & $2.31^{+0.62}_{-0.39}$ & $(3.31,3.37,3.50)$ & $2.88^{+0.25}_{-0.23}$ & $2.28^{+0.61}_{-0.61}$ \\
RXJ2129 & 
$3.18^{+0.04}_{-0.05}$ & $2.37^{+0.18}_{-0.18}$ & $3.18^{+0.05}_{-0.05}$ & $2.45^{+0.20}_{-0.18}$ & $1.50^{+0.50}_{-0.53}$ & $(1.90,2.17,2.63)$ & $2.55^{+0.26}_{-0.25}$ & $2.17^{+0.72}_{-0.72}$ \\
A611 & 
$3.32^{+0.05}_{-0.05}$ & $2.69^{+0.18}_{-0.19}$ & $3.32^{+0.05}_{-0.05}$ & $2.72^{+0.19}_{-0.18}$ & $1.32^{+0.53}_{-1.34}$ & $3.30^{+0.08}_{-0.08}$ & $3.54^{+0.22}_{-0.23}$ & $(2.18,2.27,2.74)$ \\
MS2137 & 
$3.27^{+0.08}_{-0.09}$ & $2.73^{+0.36}_{-0.38}$ & $3.27^{+0.07}_{-0.08}$ & $2.76^{+0.33}_{-0.35}$ & $0.50^{+1.21}_{-2.49}$ & $3.26^{+0.10}_{-0.09}$ & $2.88^{+0.39}_{-0.38}$ & $(1.85,2.30,2.43)$ \\
RXJ2248 & 
$3.30^{+0.06}_{-0.06}$ & $2.64^{+0.25}_{-0.23}$ & $3.31^{+0.06}_{-0.07}$ & $2.70^{+0.27}_{-0.28}$ & $-0.17^{+1.16}_{-1.73}$ & $3.28^{+0.09}_{-0.08}$ & $2.77^{+0.29}_{-0.28}$ & $(1.97,2.29,2.55)$ \\
MACSJ1115  &
$3.32^{+0.04}_{-0.04}$ & $2.85^{+0.16}_{-0.16}$ & $3.31^{+0.04}_{-0.05}$ & $2.89^{+0.17}_{-0.17}$ & $1.32^{+1.09}_{-3.06}$ & $(2.69,2.89,3.21)$ & $(3.03,3.28,3.57)$ & $(2.73,3.38,3.42)$ \\
MACSJ1931  & 
$3.30^{+0.07}_{-0.07}$ & $2.61^{+0.28}_{-0.29}$ & $3.31^{+0.07}_{-0.07}$ & $2.76^{+0.28}_{-0.24}$ & $1.49^{+0.81}_{-2.44}$ & $3.29^{+0.14}_{-0.14}$ & $2.75^{+0.32}_{-0.31}$ & $(2.09,2.21,2.80)$ \\
MACSJ1720 & 
$3.27^{+0.05}_{-0.05}$ & $2.57^{+0.18}_{-0.19}$ & $3.23^{+0.06}_{-0.08}$ & $2.88^{+0.14}_{-0.15}$ & $2.56^{+0.19}_{-0.16}$ & $3.06^{+0.15}_{-0.16}$ & $2.95^{+0.26}_{-0.26}$ & $(2.52,2.92,3.01)$ \\
MACSJ0416 & 
$3.25^{+0.04}_{-0.04}$ & $2.75^{+0.14}_{-0.14}$ & $3.24^{+0.04}_{-0.05}$ & $2.74^{+0.14}_{-0.21}$ & $-1.89^{+1.47}_{-3.43}$ & $2.65^{+0.45}_{-0.13}$ & $3.18^{+0.30}_{-0.30}$ & $(3.06,3.42,3.44)$ \\
MACSJ0429 & 
$3.21^{+0.05}_{-0.06}$ & $2.45^{+0.21}_{-0.22}$ & $3.16^{+0.08}_{-0.10}$ & $2.82^{+0.16}_{-0.16}$ & $2.56^{+0.15}_{-0.21}$ & $2.95^{+0.19}_{-0.18}$ & $2.85^{+0.26}_{-0.25}$ & $(2.58,2.59,2.96)$ \\
MACSJ1206 &
$3.29^{+0.04}_{-0.04}$ & $2.61^{+0.18}_{-0.19}$ & $3.29^{+0.04}_{-0.05}$ & $2.64^{+0.20}_{-0.19}$ & $0.63^{+1.16}_{-1.23}$ & $3.27^{+0.08}_{-0.08}$ & $2.77^{+0.25}_{-0.24}$ & $(2.26,2.32,2.79)$ \\
MACSJ0329 & 
$3.19^{+0.04}_{-0.04}$ & $2.26^{+0.15}_{-0.15}$ & $3.18^{+0.04}_{-0.05}$ & $2.34^{+0.19}_{-0.16}$ & $1.29^{+0.82}_{-2.30}$ & $(2.50,2.70,2.98)$ & $2.77^{+0.40}_{-0.38}$ & $(2.95,3.04,3.18)$ \\
RXJ1347 & 
$3.41^{+0.04}_{-0.05}$ & $2.91^{+0.17}_{-0.17}$ & $3.40^{+0.04}_{-0.06}$ & $2.96^{+0.18}_{-0.18}$ & $1.62^{+0.74}_{-0.53}$ & $3.19^{+0.19}_{-0.17}$ & $3.25^{+0.31}_{-0.32}$ & $(2.79,3.16,3.43)$ \\
MACSJ1149 & 
$3.32^{+0.04}_{-0.05}$ & $2.91^{+0.17}_{-0.17}$ & $3.32^{+0.04}_{-0.06}$ & $2.98^{+0.17}_{-0.18}$ & $0.83^{+1.61}_{-0.52}$ & $3.05^{+0.20}_{-0.18}$ & $3.29^{+0.37}_{-0.36}$ & $(2.01,3.22,3.23)$ \\
MACSJ0717 & 
$3.37^{+0.03}_{-0.04}$ & $3.12^{+0.12}_{-0.12}$ & $3.36^{+0.04}_{-0.06}$ & $3.18^{+0.14}_{-0.13}$ & $1.09^{+1.69}_{-0.94}$ & $(2.67,2.80,3.17)$ & $3.49^{+0.24}_{-0.26}$ & $3.62^{+0.71}_{-0.82}$ \\
MACSJ0647 & 
$3.26^{+0.05}_{-0.05}$ & $2.59^{+0.21}_{-0.21}$ & $3.25^{+0.05}_{-0.05}$ & $2.66^{+0.22}_{-0.20}$ & $1.50^{+0.75}_{-0.82}$ & $3.23^{+0.15}_{-0.14}$ & $2.77^{+0.29}_{-0.28}$ & $>2.46$ \\
MACSJ0744 & 
$3.25^{+0.04}_{-0.05}$ & $2.59^{+0.19}_{-0.20}$ & $3.26^{+0.04}_{-0.05}$ & $2.69^{+0.22}_{-0.15}$ & $1.18^{+0.73}_{-0.87}$ & $(2.29,2.45,3.03)$ & $3.11^{+0.40}_{-0.40}$ & $(2.97,3.54,3.62)$ \\
\hline
\hline
\end{tabular}}
\end{minipage}
\end{table*}}}

\bibliographystyle{apsrev4-1}
\bibliography{bibliography.bib}

\end{document}